# Electron-Hole Binding Governs Carrier Transport in Halide Perovskite Nanocrystal Thin Films

*Michael F. Lichtenegger\*, Jan Drewniok, Andreas Bornschlegl, Carola Lampe, Andreas Singldinger, Nina A. Henke, Alexander S. Urban\**

Nanospectroscopy Group and Center for Nanoscience (CeNS), Nano-Institute Munich, Department of Physics, Ludwig-Maximilians-Universität München,

Königinstr. 10, 80539 Munich, Germany

AUTHOR INFORMATION

**Corresponding Authors**

*mic.lichtenegger@physik.uni-muenchen.de (M.F.L.)

*urban@lmu.de (A.S.U.)

ABSTRACT

Two-dimensional halide perovskite nanoplatelets (NPLs) have exceptional light-emitting properties, including wide spectral tunability, ultrafast radiative decays, high quantum yields (QY), and oriented emission. To realize efficient devices, it is imperative to understand how exciton transport progresses in NPL thin films. Due to the high binding energies of electron-hole pairs, excitons are generally considered the dominant species responsible for carrier transfer. We employ spatially and temporally resolved optical microscopy to map exciton diffusion in perovskite nanocrystal (NC) thin films between $15\,°C$ and $50\,°C$. At room temperature (RT), we find the diffusion length to be inversely correlated to the thickness of the nanocrystals (NCs). With increasing temperatures, exciton diffusion declines for all NC films, but at different rates. This leads to specific temperature turnover points, at which thinner NPLs exhibit higher diffusion lengths. We attribute this anomalous diffusion behavior to the coexistence of excitons and free electron hole-pairs inside the individual NCs within our temperature range. The organic ligand shell surrounding the NCs prevents charge transfer. Accordingly, any time an electron-hole pair spends in the unbound state reduces the FRET-mediated inter-NC transfer rates and consequently the overall diffusion. These results clarify how exciton diffusion progresses in strongly confined halide perovskite NC films, emphasizing critical considerations for optoelectronic devices.

TOC Graphic

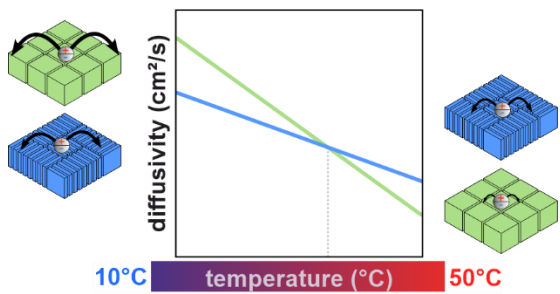

Commuting fossil fuel-relying energy production to renewable sources is necessary to combat global climate change. However, conserving energy through advanced technologies may be equally important, as witnessed in the case of the semiconductor-based light-emitting diode (LED), which rendered incandescent lightbulbs virtually obsolete.[1] Researchers and engineers have turned toward semiconductor NCs with size-dependent optical properties due to quantum confinement to further improve this technology, especially for display applications.[2, 3] One of the newer materials to have emerged for such functions, halide perovskite NCs have significant advantages over other traditional semiconductor materials.[4] The combination of quantum confinement and halide composition provides a spectral tunability from $390\ nm$ up to $800\ nm$,[5, 6] and quantum yields approaching unity have been reported.[7, 8] For various reasons, not all spectral regimes have improved in lock-step, with especially the deep blue regions lagging behind.[9] A workaround comes in the form of two-dimensional NPLs, which can be synthesized with a monolayer-precise control over thickness,[10] leading to a highly tunable, efficient blue luminescence.[11-14] Additionally, NPLs exhibit extremely-fast radiative decay rates, narrow emission linewidths, and directional emission.[15-17] Both the fundamental material properties and the NPL interaction in an ensemble are of great importance for device operation. In NPLs, high binding energies of electron-hole pairs render excitons the dominant carrier species. It is thus essential to understand how transport of excitons occurs to control and optimize it, if possible. For NPLs, knowledge on these matters is significantly lacking.[18]

Here, we present a detailed study on exciton transport in halide perovskite NPL thin films. We employ a previously developed method based on time-resolved confocal microscopy.[19] We find considerable exciton diffusivities on the order of $0.2\ \frac{cm^2}{s}$ at room temperature (RT), with the largest values in the weakly-confined nanocubes. Interestingly, we find that the diffusion becomes worse with increasing temperatures. This drop-off is thickness dependent; accordingly, as the temperature increases, first the thicker and then the thinner NPLs exhibit the highest exciton diffusion. We attribute this to the differences in exciton binding energy and a shifted equilibrium from the excitonic to the unbound state of the electron-hole pairs. This reduces the FRET rate between adjacent NCs and causes a strongly subdiffusive behavior. These findings are of extreme importance as they show that the device's operating temperature will critically influence the transport properties and must be taken into consideration during design and fabrication. This study adds another point to the list of advantages of halide perovskite NPLs for light-emitting and display applications.

RESULTS/DISCUSSION

We synthesized the halide perovskite NCs according to previously detailed syntheses.[14, 15] The three samples consist of the same material, cesium lead bromide, the exact composition of which ($Cs_{n-1}Pb_nBr_{3n+1}$) varies based on the thickness of the resulting NCs. Two samples comprise two-dimensional NPLs, with $n = 3$ and $n = 5$ monolayer (ML) thickness and one sample of squarish nanocubes. Due to strong confinement in one dimension, the NPLs exhibit a thickness-dependent, strongly blue-shifted photoluminescence (PL) emission compared to the weakly-confined nanocubes (Figure 1a). Transmission electron microscopy (TEM) reveals the cubes to have a side length of $14\ nm$ and the NPLs to have lateral dimensions of $20\ nm$ and thickness of $1.8 \pm 0.2\ nm$ (3ML) and $3.0 \pm 0.2\ nm$ (5ML) (see Supporting Information (SI) Figure S1). The NCs are stabilized with oleylamine and oleic acid ligands, leading to an organic separation layer of $2.2 \pm 0.5\ nm$ between adjacent NCs. These spacings are too large to facilitate charge transfer via tunneling between the NCs.[19, 20] Accordingly, any experimentally determined macroscopic transport must occur *via* energy transfer. The phenomenological theory for this nonradiative process is known as Förster resonance energy transfer (FRET). [21, 22] The efficiency of this process, which occurs *via* dipole-dipole coupling, is highly distance-dependent with a sixth-power law for two point dipoles: $\eta_{FRET} = \frac{1}{1+(r/R_0)^6}$. Therein, $R_0$ is the so-called Förster radius between the donor-acceptor pair and signifies the distance at which the energy transfer efficiency amounts to 50 %. The Förster radius itself depends strongly on the overlap integral of the donor emission spectrum with the acceptor absorption spectrum and the orientation of the two transition dipole moments. This orientation is maximized in the case of NPLs aligning in parallel, and can enable transfer efficiencies of up to 70 % at comparable NPL spacings.[23-25]

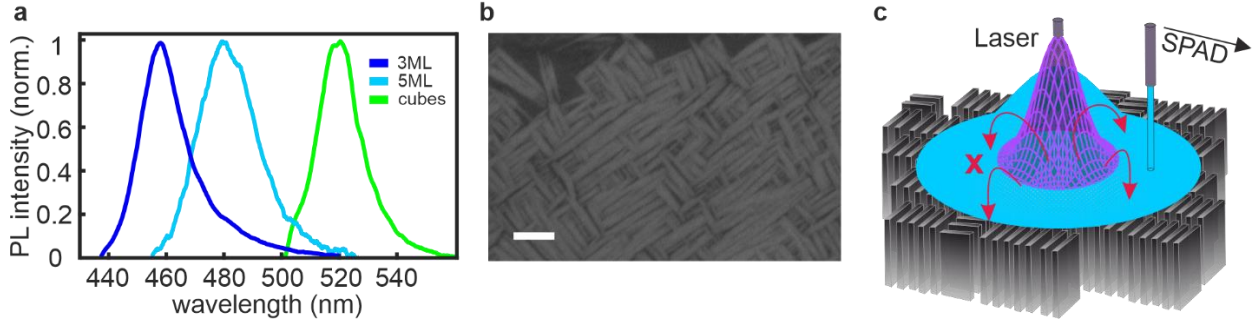

**Figure 1.** (a) PL spectra of perovskite NC films consisting of 3ML NPLs (dark blue), 5ML NPLs (light blue) and cubes (green). The PL prominently blueshifts with decreasing thickness of the NCs. (b) SEM image of a 3ML NPL film deposited on SiO$_2$-coated Si substrates (scale bar is 400 $nm$). Upon drying, the NPLs align face-on to form stacks up to micrometers in length. The stacks themselves align in parallel in small bundles or oriented orthogonally. (c) Sketch of the measurement method. The perovskite NC film is excited *via* a pulsed Gaussian laser beam (violet). Excitons in excited NCs can recombine or hop to adjacent NCs, leading the PL spot broadening. A SPAD placed in the microscope's focal plane is scanned over the focal emission spot, enabling a spatial and temporal resolution of the exciton dynamics.

To measure exciton diffusion, we drop-cast the NCs dispersed in hexane onto $SiO_2$-coated Si substrates. Due to a very high homogeneity in thickness, the NPLs self-assemble in edge-up configuration with stack lengths of up to a few micrometers in length as can be determined by scanning electron microscopy (SEM) (Figure 1b shows a 3ML NPL film and SI Figure S2 5ML NPL and nanocube films). This was also seen previously for CdSe NPLs.[26-28] TEM images show these stacks in detail, with individual NPLs, their orientation, and the spacings between them distinguishable (SI Figure S1). The individual stacks tend to arrange in parallel bundles, with these bundles aligning roughly orthogonally to each other. The nanocubes form relatively dense films with high order only on short length scales and some voids (typically $< 100\ nm$ in size) visible. Film thicknesses of $100 - 200\ nm$ are determined by SEM images (SI Figure S3). The diffusion measurements were carried out in a home-built time-resolved confocal microscope based upon the method introduced by Akselrod *et al.*.[19] Briefly, as sketched in Figure 1c, a narrow spectral band of a white light laser with picosecond pulses is focused to a diffraction-limited Gaussian spot on the NC film. The light is partially absorbed in the film, creating electron-hole pairs, which rapidly relax to form tightly bound excitons.[29] Importantly, we excite at very low laser fluence to ensure that at most one exciton is created per individual NC. The initial exciton distribution in the film is given by the absorption of the excitation laser and can be described by a Lorentzian profile.[19, 30] A consequence of this is a strong concentration gradient within the NC film, which promotes exciton diffusion. The PL lifetime of the NCs lies in the nanosecond range (SI Figure S4), meaning that excitons recombine (radiatively and nonradiatively) within a few nanoseconds. During this time, they can diffuse from their point of origin. Accordingly, for $t > 0$, the distribution of emitted photons should differ from the initial exciton distribution. This can be monitored by imaging the PL on the sample, which shows a slightly larger spot size (due to the integrated diffusion process). This, however, only reveals the steady-state case and does not allow any insight into exciton dynamics. In order to obtain this information, the PL is collected *via* an optical fiber positioned in the focal plane and passed to a single photon avalanche detector (SPAD) with a time resolution of $< 50\ ps$. The fiber is scanned over the film's surface with step sizes of at least $40\ nm$ near the excitation spot and slightly larger ones further away from the center. Thus, we obtain PL emission of the entire sample with high temporal and spatial resolution. (It is important to note that we are not diffraction limited by this method, as we are not resolving different spots, instead only monitoring local changes.) The PL data is then combined to form a two-dimensional graph depicting the PL intensity over the displacement from the central excitation spot and over time after excitation (Figure 2a). The PL map is shown for a 3ML NPL film at RT. The PL intensity, normalized to each time slice ($25\ ps$ step size), can be seen to become slightly broader within the $3.5\ ns$ shown here. This broadening is a clear indication for exciton migration in the NC films. The broadening can be seen more easily by extracting the spatial PL profiles at given times (marked with dashed lines in Figure 2a) and comparing these to each other (Figure 2b). Here, the ($25\ ps$ wide) slices at $0, 1, 2,$ and $3\ ns$, shown from dark to light blue, display a progressive broadening. To quantify this broadening, we fit the spatially resolved PL with a Voigt profile, which is the convolution of a Gaussian and Lorentzian profile and best describes the initial exciton population immediately after excitation (SI Figure S5).[19] The exciton diffusion leads to a broadening of the profile, given by the standard deviation of the Gaussian profile, $\sigma$. To quantify this broadening, we determine the mean squared displacement (MSD) of the exciton population, given by $MSD(t) = \Delta\sigma^2 = \sigma(t)^2 - \sigma(0)^2$.[30] The initial width is given by the first time slice of the PL measurements at 25 $ps$ and is by definition $MSD(0) = 0$. We thus fit a Voigt profile to each time slice and determine the Gaussian standard deviation for each. As the excitons population decays over time and the signal to noise ratio worsens, it becomes more challenging to determine the distribution. Therefore, we terminate the fitting once the relative error of the Voigt profile exceeds 10 %. For the NPLs, the cutoff time is typically after $5 - 6\ ns$, at which only about 15 % of the initial exciton population still exists. For the nanocubes, the range extends out slightly farther, to approximately 10 $ns$, since nanocubes are commonly brighter than NPL films due to higher QYs. The resulting MSDs at RT are depicted for the three different NC films in Figure 2c. The MSD of the nanocubes exhibits an apparent sublinear increase with time and is significantly larger than for the two NPL samples. Typical diffusive processes (Brownian motion) in undisturbed systems (*e.g.*, no traps) are characterized by a linear temporal increase of the $MSD(t) = \Delta\sigma^2 = 2d \cdot Dt$ with the dimension of the system, $d$ and the constant diffusion coefficient or diffusivity, $D$. Anomalous diffusion, in contrast, follows a power law according to $MSD(t) = \Delta\sigma^2 = K_\alpha \cdot t^\alpha$ with the generalized diffusion coefficient $K_\alpha$. Subdiffusion is characterized by $\alpha < 1$ and is a sign of restricted diffusion through crystal boundaries, voids, inhomogeneous energy landscapes and traps.[19, 31] An alternative method to describe the diffusion in 2D perovskites is to involve traps, which bind excitons and prevent their further diffusion.[30, 32] Applying this model to our data also yields noticeable results, especially for the NPL films, with similar diffusion parameters. Accordingly, we merge all these mechanisms, employing the subdiffusive model with two free parameters as in Eq.(1). The diffusion behavior is well characterized by the three parameters: exciton lifetime $\tau$, diffusivity $D$, and diffusion length $L_D$. For the subdiffusive behavior, the diffusivity is not constant but time-dependent. It can be determined, using the fit parameter $K_\alpha$ through $D(t) = \frac{1}{2}\frac{d\,MSD(t)}{dt} = \frac{1}{2}K_\alpha \cdot \alpha \cdot t^{\alpha-1}$. While the overall MSD decreases with decreasing NC thickness, the parameter $\alpha$ displays the opposite trend, approaching 1 for the 3ML NPLs. The resulting averaged diffusivity $\bar{D}$ (see SI) of the nanocubes amounts to $0.32\,\frac{cm^2}{s}$ and is significantly larger than for the 3ML NPLs ($0.1\,\frac{cm^2}{s}$) but not much higher than for the 5ML NPLs ($0.28\,\frac{cm^2}{s}$). The PL spectrum of the 3ML NPLs is slightly asymmetric with a long tail towards the low-energy side due to trap states, which we confirm with PL measurements of the NPLs in dispersion and at low temperatures in films (SI Figure S6). To rule out that these trap states are the cause of the much lower diffusivity, we add spectrally resolved diffusion measurements. For this, we insert a short-pass filter into the detection path and cut off the PL emission above $465\ nm$ (SI Figure S7). With no appreciable difference to the diffusion obtained for the full spectrum, we can rule out the inhomogeneously distributed traps as a cause of the low diffusivities.[19] Interestingly, we find that the diffusion does not depend on excitation density for the two NPL samples but increases significantly for higher laser fluences for the nanocubes (SI Figure S8). It is important to note that the low excitation density ensures that we are still in the linear excitation regime (which we confirm with Poisson statistics, see SI Tables S3 and S4) and precludes the occurrence of exciton-exciton annihilation (as in previous studies, SI Figure S9).[31]

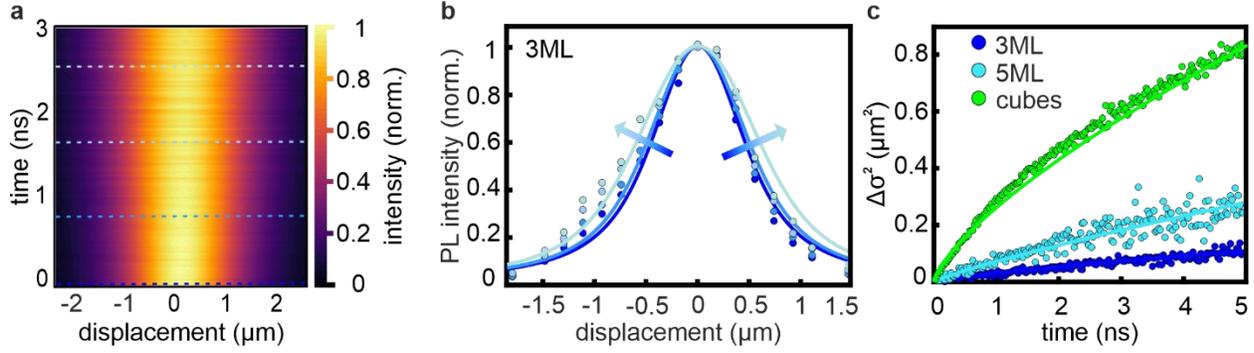

**Figure 2. Mapping exciton diffusion in perovskite NC films.** (a) Map of the temporal evolution of the PL intensity in a 3ML NPL film at RT. (b) Cross-sections of the experimental PL map in (a) are taken at times denoted by the dashed lines (from 0 to 3 $ns$, colors are matched). The data are fitted with Voigt functions (solid lines), showing a profile broadening with time. (d) This broadening can be quantified by the MSD shown here for 3ML NPLs (dark blue), 5ML NPLs (light blue), and nanocubes (green). The experimental data points (dots) are fitted with the subdiffusive model described in the main text.

Both the nanocubes and NPLs have advantages and disadvantages for energy transport. As the nanocubes are significantly thicker, the average distance traveled each time the exciton jumps between adjacent NCs is larger than the NPLs. Thus, assuming an equal attempt rate (or exciton hopping rate) for a FRET process, excitons will diffuse further in nanocube films. The hopping rate itself is given by the equation: $\tau_{DA}^{-1} = k_{DA} = \frac{2\pi}{\hbar}\frac{\mu_D^2\mu_A^2\kappa^2}{r_{DA}^6 n^4}$ with the interparticle distance $r_{DA}$, the refractive index $n$, the spectral overlap between donor and acceptor $\mu_D^2\mu_A^2$, and the dipole alignment factor $\kappa^2$.[31] This equation highlights several advantages for the NPLs: a lower effective refractive index,[17] a more extensive spectral overlap due to more negligible inhomogeneous broadening, and, most importantly, a more favorable orientation of the transition dipoles. In nanocubes, these are reasonably isotropic, while in NPLs, they are significantly oriented.[16,17] Additionally, the exciton Bohr radius in NPLs is larger than the width of the inorganic layers, extending into the surrounding ligands. Together with the propensity of the NPLs to form stacks, wherein the dipoles are strongly aligned, these factors will likely lead to a higher hopping rate than for the nanocubes films. To investigate how the different hopping rates and interparticle distances affect the overall exciton diffusion, we look at the basics of the FRET process, according to which an increased temperature should enhance the hopping rate.[9, 33] Therefore, we constructed a simple stage with a coupled Peltier device to control the temperature of the films and study the effect on the diffusion parameters. The accessible temperature range is limited to $15 - 55\ °C$, as at lower temperatures, the moisture from the ambient air condenses on the samples, while at higher temperatures, the NC films begin to degrade irreversibly.

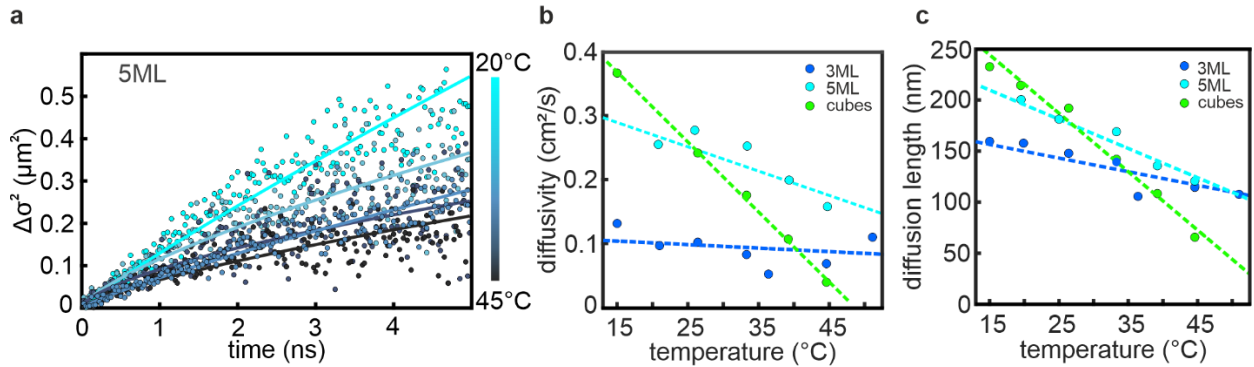

**Figure 3. Temperature-dependent exciton diffusion in perovskite NC films.** (a) Temporal MSD data (filled circles) for a 5ML NPL film in the temperature range 20 °C-45 °C. For increasing temperature, the MSD values decrease, and the diffusion becomes more subdiffusive. The solid lines correspond to fits according to (1). (b) Diffusivity as a function of temperature for each perovskite NC film under consideration. The dashed lines are guides to the eye. (c) Diffusion length as a function of temperature for each perovskite NC film under consideration. The dashed lines are guides to the eye.

Temperature-dependent MSD curves are shown for the 5ML NPLs in Figure 3a (see SI for data on the other two samples - Figure S10). As the temperature is increased, the MSD and concomitantly the diffusion decreases. We model the data with the subdiffusive equation mentioned above and find that the process becomes significantly more subdiffusive, with $\alpha$ decreasing from $0.9\ to\ 0.74$ (for nanocubes $0.89\ to\ 0.66$ and 3ML NPLs $0.89\ to\ 0.87$). We extract the diffusivity from the MSD data at RT and obtain a value of $0.28\ \frac{cm^2}{s}$, which fits right in between the determined value for the 3ML NPLs $\left(0.1\ \frac{cm^2}{s}\right)$ and the nanocubes $\left(0.31\ \frac{cm^2}{s}\right)$. These are significantly larger than those measured in typical quantum dot films and comparable to previous results on halide perovskite NC films.[19, 31, 34-36] Interestingly, we find a negative linear correlation with temperature $(D = -A \cdot T)$ and $A = 3.3 \cdot 10^{-3} \frac{cm^2}{s \cdot K}$ (Figure 3b). The same trend holds for both the thinner, 3ML NPLs and the nanocubes: the diffusivity decreases progressively with temperature $\left(\frac{\partial D}{\partial T} < 0\right)$. Intriguingly, the slope depends strongly on the NC thickness, with the diffusivity decreasing more strongly with temperature in films of thicker NCs. Accordingly, while the nanocubes exhibit the highest diffusivity at RT, for $T \approx 26 - 72\ °C$ the 5ML NPLs exhibit the highest diffusivity. At $40\ °C$ the diffusivity of the nanocubes drops below that of the 3ML NPLs, and if we extrapolate out to higher temperatures at $72\ °C$ the 3ML NPLs would display the highest diffusivity. This behavior becomes even more significant when comparing the diffusion lengths in the three NC films (Figure 3c). The diffusion length is defined as $L_D = \sqrt{\tau \overline{D}}$ and describes how far an exciton can diffuse on average until it recombines. Qualitatively, the diffusion length follows the same trend as the diffusivity. While the nonradiative decay rate increases for higher temperatures, the exciton decay rate decreases only insignificantly (SI, Figure S11). As for the diffusivities, the diffusion lengths are highest in the nanocubes at RT ($220\ nm$), followed by the 5ML NPLs ($200\ nm$) and then the 3ML NPLs ($155\ nm$). The value for the nanocubes matches previously determined values.[9, 31, 35] The crossover points are slightly shifted to lower temperatures, with the diffusion length of the nanocubes dipping below that of the 3ML NPLs at 35 °$C$ and the 3ML NPLs exhibiting the largest diffusion lengths above 50 °$C$. These trends seem surprising initially, as increased temperatures should make the hopping easier, enhancing the diffusion. Instead, we observe the opposite behavior, reminiscent of band-like transport.[30] However, with a spacing between the NPLs of roughly 2 $nm$, we can safely rule this out as an explanation.[37] Additionally, we do not observe a progressive redshift of the PL emission with increasing times after excitation or an emission wavelength-dependent decay rate (SI Figure S11). Both of these phenomena would point towards a substantial downhill energy transfer process.[19, 32] We can also rule out polarons as a cause for the observed transport properties. Small polarons located at a single lattice site also exhibit an increased diffusivity with temperature, opposite to our observations. Large polarons, on the other hand, show the same temperature dependence ($\partial D/\partial T < 0$), however, they are unlikely to transfer between NPLs, whose lateral dimensions are only 14 $nm$.[32]

One essential result of the strong confinement in the NPLs is the drastic increase in exciton binding energies compared to bulk values. While it is generally assumed that excitons are prevalent once the binding energy is larger than the thermal energy, this is not strictly the case. In bulk films, excitons can dissociate despite this larger binding energy and provide charge transport.[38, 39] However, even for nanocrystals, the electrons and holes transition between the bound (exciton) and unbound (free carriers) states, with the equilibrium shifting toward the bound state as the binding energy increases.[40] The temperature strongly affects the equilibrium between bound and unbound states, especially for binding energies comparable to the thermal energy. The effect of this can be estimated by using Saha's equation.[41] Using previously reported values for the binding energies,[14] we find that in the temperature range investigated here, the fraction of excitons to all photogenerated electron-hole pairs depends strongly on the thickness of the NCs (SI Figure S12). Additionally, there is an obvious temperature-dependence in this range, with the exciton fraction declining steeply for the nanocubes but less steeply for the NPLs. This is an extremely important finding, as in the form of free charge carriers, the intercrystal mobility drops to nearly zero due to vastly diminished tunneling probabilities at separations of more than 1 nm.[42] This explains the grade of the slopes in the diffusivity curves, with the nanocubes ($E_B \approx 35\ meV$) being most strongly affected and the 3ML NPLs ($E_B \approx 180\ meV$) the least. We also find that the exponent $\alpha$ describing the subdiffusive behavior follows the same trend. While $\alpha$ becomes progressively smaller with increasing temperature, it does so much more rapidly the thicker the NCs are. We can thus surmise that the formation and dissociation of excitons strongly affect the diffusion process. Effectively, as diffusion is non-existent for free charge carriers, the subdiffusive nature increases with the fraction of time that electron-hole pairs exist in the unbound state. These findings also help to explain the strong fluence-dependence of the diffusivities in the nanocubes. The free electrons and holes are far more susceptible to trapping than excitons. Higher excitation densities consequently lead to more electrons and holes passivating traps, enhancing the diffusivity for subsequent excitons. We confirm this by additionally performing the diffusion measurements with an concomitant contiuous LED illumination at $365\ nm$ (SI Figure S13). Increasing the incident LED light leads to a substantial increase in the diffusivity so that it become comparable to diffusion at much higher excitation densities. Summarily, at lower temperatures (at or below RT), the nanocube films have better exciton transport properties, while at slightly elevated temperatures, the NPL films prove more advantageous.

CONCLUSIONS

In conclusion, we have investigated exciton diffusion in thickness-controlled halide perovskite NC films *via* transient confocal microscopy. The process is subdiffusive ($\alpha < 1$) with diffusivities $\left(D = 0.1 - 0.3 \frac{cm^2}{s}\right)$ and diffusion lengths ($L_D = 150 - 220 \, nm$) significantly larger than in conventional QD films and comparable to previous studies on halide perovskite NCs. The diffusion properties depend on the thickness of the NCs, with the nanocubes exhibiting the highest values at RT and the 3ML NPLs the lowest values. Employing a Peltier stage to control the temperature in the NC films, we find that these properties generally decrease with increasing temperatures. Notably, the decrease is most substantial for the thickest NCs, and so above 40 °$C$ the 5ML NPLs have the largest diffusion, and at even higher temperatures, the 3MLs exhibit the largest values. We conclude that the diffusion process is FRET-mediated with excitons hopping between adjacent NCs. The exciton binding energy is the critical parameter in the process, as there is an equilibrium between exciton formation and dissociation. The organic ligand spacers surrounding the NCs prevent tunneling of charge carriers. Thus free charge charrier diffusion only occurs within individual NCs and not between NCs. The equilibrium is shifted toward the unbound state for higher temperatures, further limiting diffusion and most strongly affecting the nanocubes, whose exciton binding energies are slightly above the thermal energy at RT ($E_B \approx 35 \, meV$). Consequently, at even mildly elevated temperature ($T > 35°C$) the NPLs exhibit better diffusive properties. These results highlight another advantage of NPLs over typical isotropic NCs for optoelectronic devices, in addition to improved quantum yields in the blue spectral range and an enhanced outcoupling. Thus, research should focus on further enhancing the PL quantum yields and improving the stability of these exciting colloidal quantum wells to obtain high performance, high-brightness LEDs and displays.

METHODS/EXPERIMENTAL

Synthesis

Nanoplatelets: $Cs_2CO_3$ (cesium carbonate, 99 %), $PbBr_2$ (lead(II) bromide, $\geq$ 98 %), oleic acid (technical grade 90 %), oleylamine (technical grade 70 %), acetone (for HPLC, $\geq$ 99.9 %), toluene (for HPLC, $\geq$ 99.9 %) and hexane (for HPLC, $\geq$ 97.0 %, GC) were purchased from Sigma-Aldrich. All chemicals were used as received. The $PbBr_2$ and Cs precursor solutions and the photoluminescence (PL) enhancement solution were prepared as reported previously.[14] The synthesis of the NPLs, carried out under an ambient atmosphere at room temperature, was modified slightly. A vial was charged with a $PbBr_2$ precursor solution, and a Cs oleate precursor solution was added under stirring at 1200 rpm. The corresponding volumes for different thicknesses are listed in Supporting Table 1. After ten seconds, acetone was added, and the reaction mixture was stirred for one minute. The mixture was centrifuged at 4000 rpm for 3 min and the precipitate was redispersed in hexane (1.8 ml). A PL enhancement solution (200 µl) was added subsequently to enhance the stability and emission properties of the NPLs.

Nanocubes: The nanocubes were synthesized as previously established.[43]

NC films were prepared by dropcasting 40 µl of the NC dispersion onto a $SiO_2$-coated (300 nm) 10 × 10 mm Si substrate.

Microscopy Methods:

Steady-state and time-resolved confocal microscopy measurements were conducted using a pulsed laser (NKT Photonics, SuperK Fianium FIU-15) and a tunable multi-line filter (SuperK SELECT) to select an excitation wavelength depending on the specific NC according to Supporting Table 2. The laser was operated with a repetition rate of $1.95\ MHz$ and afluence of $8\ \mu J/cm^2$ arriving at the sample. This fluence ensures a low exciton generation probability per NC (around 1-5 %), minimizing the influence of higher-order processes like Auger recombination or exciton-exciton annihilation. The PL was collected with an objective (Nikon, Tu Plan Apo EPI 100x NA 0.9) with a working distance of $2.0\ mm$. The PL signal was separated from the excitation laser with suitable dichroic beam splitters for the three NCs: Semrock, Beamsplitter HC BS 495, Chroma, Beamsplitter T 450 LPXR, and Chroma, Beamsplitter T 425 LPXR for the nanocube, 5ML NPL and 3ML NPL films, respectively. Steady-state PL spectra were obtained using a fiber-coupled (fiber diameter $200\ \mu m$) portable spectrometer (Thorlabs, Compact CCD Spectrometer). The fiber is mounted within the microscope focal plane on top of a 3-axis stage (Thorlabs, Three MT1-Z8 Single-Axis). Time-resolved PL measurements were conducted using a single-photon avalanche diode (SPAD) (Micro Photon Devices, PDM series). The SPAD is fiber-coupled and mounted on the aforementioned stage, enabling the spatially resolved acquisition of decay traces within the focal plane. A fiber diameter of $10\ \mu m$ was chosen, leading to a microscope resolution of $40\ nm$. Time-correlated single-photon counting (TCSPC) measurements were realized by a TCSPC-card (PicoQuant, TimeHarp 260). A Peltier element (QuickCool, QC-127-1.4-8.5MD) beneath the Si substrate is used to set the temperature in the NC film. The temperature range is thereby limited to $15\ °C - 50\ °C$. Below $15\ °C$ water vapor from the ambient air condenses on the samples, damaging the perovskite NCs. Temperature above $55\ °C$ lead to an irreversible decrease of the NC PL, likely due to degradation of the perovskite NCs. The Peltier element turns on and off periodically to maintain the desired temperature. This results in a minor spatial oscillation of the substrate (duration of approximately five seconds) along the optical axis around the working distance of the objective. Since the oscillation duration is far less than the actual TCSPC measurement duration of one minute, this focusing/defocusing process is averaged out.

Spectrally resolved TCSPC was conducted using a femtosecond fluorescence up-conversion setup (Lightconversion, Harpia TF). A Yb:YAG laser (Lightconversion, Carbide $40\ W$) was coupled into an optical parametric amplifier (Lightconversion, Orpheus HF) where the pump pulse ($\lambda_{exc} = 400\ nm$ for 3ML NPLs, $\lambda_{exc} = 465\ nm$ for nanocubes) was generated. This pulse was focused onto a $40\ \mu m$ spot on the sample. The sample fluorescence was guided into a spectrograph (Oxford Instruments, Andor Kymera 193i) with a photomultiplier tube connected to a TCSPC module (Becker & Hickl, SPC 130 EM). The spectrally resolved TCSPC was acquired with a step size of $0.5\ nm$ (3ML NPLs) and $0.3\ nm$ (nanocubes).

Transmission electron microscopy (TEM) images were recorded using a JEOL JEM-1011 with an accelerating voltage of $80\ kV$.

Scanning electron microscopy (SEM) images were recorded using a ZEISS SEM Ultra Plus with an acceleration voltage of $3\ kV$.

ASSOCIATED CONTENT

**Supporting Information.** Data analysis and diffusion models, TEM and SEM characterization and additional diffusion microscopy data. The following files are available free of charge.

Data Analysis, Supporting Figures (PDF)

AUTHOR INFORMATION

**Corresponding authors**

*mic.lichtenegger@physik.uni-muenchen.de (M.F.L.)

*urban@lmu.de (A.S.U.)

**Author Contributions**

The manuscript was written through contributions of all authors. All authors have given approval to the final version of the manuscript.

ACKNOWLEDGMENT

This project was funded by the European Research Council Horizon 2020 through the ERC Grant Agreement PINNACLE (759744), by the Deutsche Forschungsgemeinsschaft (DFG) under Germany's Excellence Strategy EXC 2089/1-390776260 and by the Bavarian State Ministry of Science, Research and Arts through the grant "Solar Technologies go Hybrid (SolTech)".